\documentclass{article}

\usepackage{PRIMEarxiv}

\usepackage[utf8]{inputenc} 
\usepackage[T1]{fontenc}    
\usepackage{hyperref}       
\usepackage{url}            
\usepackage{booktabs}       
\usepackage{amsmath}
\usepackage{amsthm}
\usepackage{amssymb}
\usepackage{amsfonts}
\usepackage{mathtools}
\usepackage{amsfonts}       
\usepackage{nicefrac}       
\usepackage{microtype}      
\usepackage{fancyhdr}       
\usepackage{graphicx}       
\usepackage{times}
\usepackage{subfig}
\usepackage{xcolor}
\usepackage{array}
\usepackage{cite}
\usepackage{multirow}
\usepackage{adjustbox}
\usepackage{fontawesome}
\usepackage{algorithm,algorithmic}
\usepackage[font=footnotesize,labelfont=bf]{caption}
\usepackage{tikz-network}
\tikzstyle{block} = [draw, fill=blue!10, rectangle, 
    minimum height=1cm, minimum width=1cm]
\tikzstyle{sum} = [draw, fill=black!20, circle, node distance=1.5cm]
\tikzstyle{input} = [coordinate]
\tikzstyle{output} = [coordinate]
\tikzstyle{pinstyle} = [pin edge={to-,thin,black}]
\usepackage{enumitem}

\title{
Neural-Enhanced Micro-Kalman Filtering for Satellite Tracking: A Comparative Study
}

\author{
  Moh Kamalul Wafi \\
  Department of Engineering Physics \\
  Institut Teknologi Sepuluh Nopember (ITS) \\
  Surabaya, Indonesia\\
  \texttt{\{kamalul.wafi\}@its.ac.id} \\
}

\begin{document}
\maketitle

\begin{abstract}
Satellite state estimation plays a fundamental role in orbital navigation, tracking, and autonomous space operations. Accurate estimation remains challenging due to uncertainties in process and measurement noise, which may degrade the performance of conventional Kalman filtering techniques. This paper presents a Neural-enhanced micro-Kalman filter ($\mu$KF) for satellite tracking based on an information-form state estimation framework. Starting from a linearized state-space model of orbital dynamics, a lightweight neural scaling mechanism is introduced to adapt the process and measurement noise covariances online while preserving the underlying Bayesian filtering structure. The proposed estimator is formulated within the information-form $\mu$KF framework and evaluated through numerical simulations using a linear Gaussian satellite tracking model. Its performance is compared with the classical Kalman filter (KF), the extended Kalman filter (EKF), the unscented Kalman filter (UKF), and an adaptive Kalman filter under identical operating conditions. Simulation results demonstrate that the proposed Neural-$\mu$KF accurately tracks the satellite states with consistently low mean square estimation errors (MSEE). Furthermore, the proposed method achieves estimation performance comparable to, and for selected states slightly better than, the baseline Kalman filter while retaining the computational advantages of the information-form formulation. These results demonstrate that integrating lightweight neural covariance adaptation into the $\mu$KF provides an effective and flexible framework for satellite state estimation.
\end{abstract}
\allowdisplaybreaks


\section{Introduction}

Satellite tracking plays a fundamental role in modern aerospace engineering, including orbital navigation, Earth observation, deep-space exploration, and communication systems. Accurate knowledge of a satellite's position and velocity is essential for trajectory prediction, collision avoidance, attitude determination, and autonomous mission planning. Since direct measurements are inevitably corrupted by sensor inaccuracies and environmental disturbances, reliable state estimation algorithms remain an indispensable component of satellite navigation systems \cite{b1,b2}.

Among the available estimation techniques, the Kalman filter has become one of the most influential methods for linear stochastic systems since its introduction by Kalman in 1960 \cite{b4}. By recursively combining a mathematical model with noisy measurements, the Kalman filter provides the minimum mean-square error estimate under linear Gaussian assumptions. Its theoretical foundation and numerous engineering applications have been extensively studied in the literature, making it a standard tool for aerospace, robotics, navigation, and signal processing problems \cite{b5,b6,b9}. Numerous improvements have subsequently been proposed to enhance the performance of the classical Kalman filter, including covariance adaptation and gradient-based optimization techniques for uncertain environments \cite{b3}.

As estimation problems become increasingly complex, several extensions of the classical Kalman filter have been proposed. For nonlinear systems, the Extended Kalman Filter (EKF) and the Unscented Kalman Filter (UKF) approximate the Bayesian filtering problem using linearization and sigma-point transformations, respectively. Adaptive Kalman filtering techniques further improve estimation performance by adjusting the process and measurement noise statistics online when these quantities are uncertain or time-varying \cite{F1a,F1b}. More recently, machine learning approaches have also been incorporated into estimation algorithms to enhance covariance adaptation and improve robustness under varying operating conditions \cite{F2a,F2b}.

An alternative formulation of the Kalman filter is the information filter, in which the estimation problem is expressed in terms of information matrices rather than covariance matrices. This representation has attracted considerable attention because of its computational advantages and its suitability for distributed estimation. Building upon this framework, Olfati-Saber introduced the micro-Kalman filter ($\mu$KF), which later became an important building block for distributed Kalman filtering in sensor networks \cite{b7,b8}. Although the information-form formulation has been extensively investigated, the incorporation of lightweight neural covariance adaptation within this framework has received comparatively less attention for satellite state estimation.

Motivated by these observations, this paper proposes a Neural-enhanced micro-Kalman filter ($\mu$KF) for satellite tracking. Starting from a linearized state-space model of orbital dynamics, a lightweight neural scaling mechanism is incorporated into the information-form filtering framework to adapt the process and measurement noise covariances online while preserving the underlying Bayesian estimation structure. The proposed estimator is evaluated through numerical simulations and compared with the classical Kalman filter (KF), the Extended Kalman Filter (EKF), the Unscented Kalman Filter (UKF), and an adaptive Kalman filter under identical operating conditions.

The main contributions of this paper are summarized as follows:
\begin{itemize}[leftmargin=*]
    \item A Neural-enhanced information-form micro-Kalman filter is proposed by introducing adaptive neural covariance scaling while preserving the original information-form filtering equations.
    \item The proposed estimator is compared with the classical Kalman filter, EKF, UKF, and an adaptive Kalman filter through extensive numerical simulations.
    \item Simulation results demonstrate that the proposed Neural-$\mu$KF achieves accurate state estimation while maintaining the computational advantages of the information-form filtering framework.
\end{itemize}

\section{Problem Formulation}
\label{sec:problem}

Consider the discrete-time linear stochastic system
\begin{subequations}
\begin{align}
    x_{k+1} &= F x_k + q_k,
    \label{eq:problem_state}\\
    y_k &= H x_k + v_k,
    \label{eq:problem_output}
\end{align}
\end{subequations}
where $x_k\in\mathbb{R}^{n}$ denotes the system state and
$y_k\in\mathbb{R}^{m}$ is the corresponding measurement at the sampling instant $k$. The matrices
$F\in\mathbb{R}^{n\times n}$ and
$H\in\mathbb{R}^{m\times n}$ denote the state-transition and measurement matrices, respectively.

The process disturbance and measurement noise are assumed to satisfy
\begin{equation}
    q_k \sim \mathcal{N}(0,\Sigma_q),
    \qquad
    v_k \sim \mathcal{N}(0,\Sigma_v).
    \label{eq:problem_noise}
\end{equation}
where $\Sigma_q\succeq0$ and $\Sigma_v\succ0$ are the corresponding covariance matrices. Furthermore,
$\mathbb E[q_kv_\ell^\top]=0,\forall k,\ell,$
that is, the process and measurement noises are mutually independent.

Given the measurement sequence
\begin{equation}
    \mathcal Y_k
    :=
    \left\{
    y_0,y_1,\ldots,y_k
    \right\},
    \label{eq:measurement_history}
\end{equation}
the objective is to determine an estimate
$\hat{x}_k$ of the unknown state $x_k$ that minimizes the mean-square estimation error. This leads to the minimum mean-square error (MMSE) estimation problem
\begin{equation}
    \hat{x}_k
    =
    \arg\min_{\hat{x}}
    \mathbb E
    \left[
        \|x_k-\hat{x}\|_2^2
        \,\middle|\,
        \mathcal Y_k
    \right].
    \label{eq:mmse_problem}
\end{equation}

This paper investigates the above estimation problem using the proposed Neural-Enhanced micro-Kalman filter ($\mu$KF). The estimation performance is evaluated through comparisons with the Extended Kalman Filter (EKF), the Unscented Kalman Filter (UKF), and an adaptive Kalman filter under identical system and measurement conditions. The mathematical model of the satellite tracking system is developed in the following section.

\section{Mathematical Model}
\label{sec:model}

\begin{figure}[t!]
    \centering
    \includegraphics[width=0.65\linewidth]{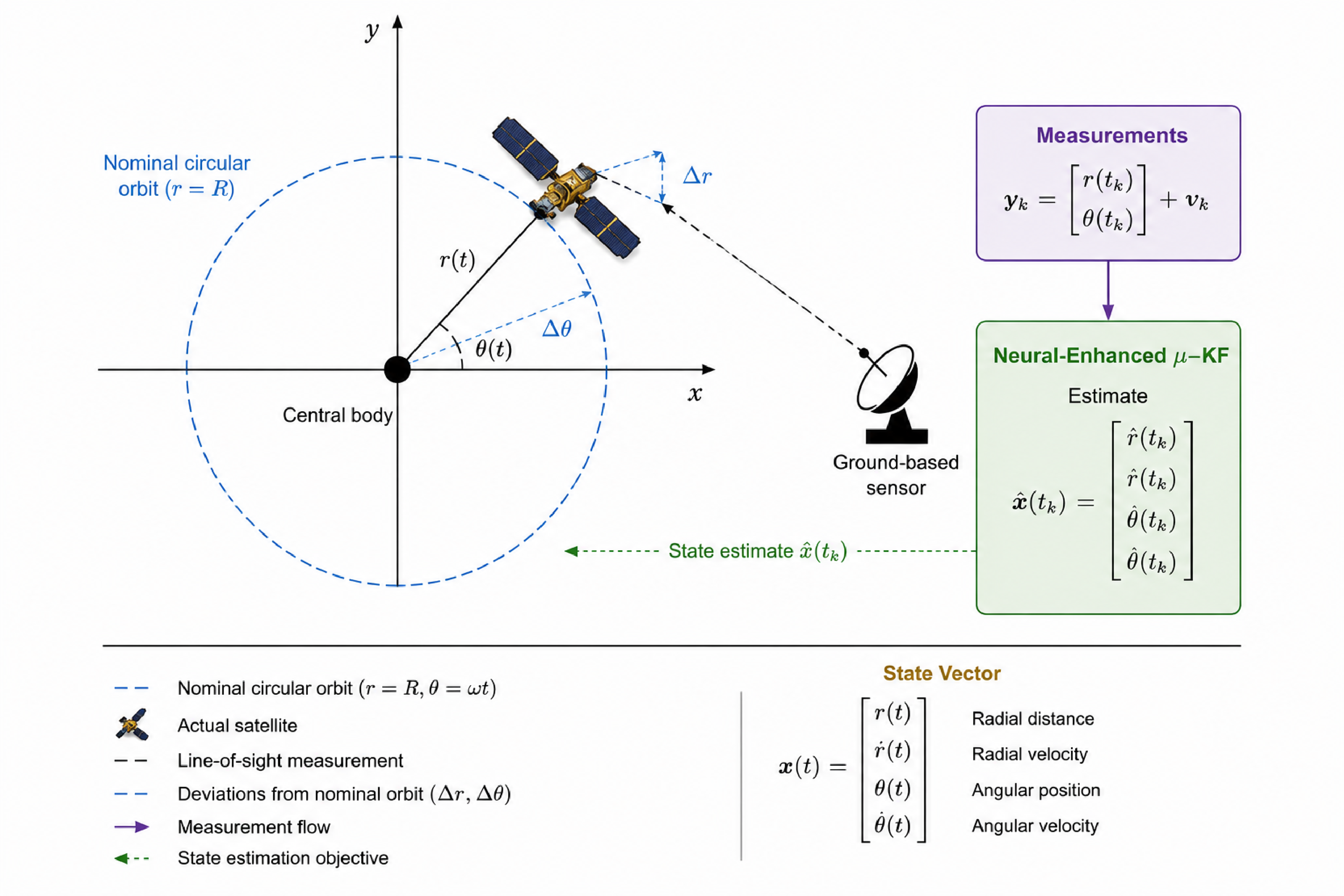}
    \caption{Block diagram of the satellite state estimation problem. The nonlinear satellite dynamics generate noisy measurements, which are processed by the proposed Neural-Enhanced micro-Kalman filter ($\mu$KF) to reconstruct the state trajectory.}
    \label{fig:problem_setup}
\end{figure}

Based on the estimation problem formulated in Section~\ref{sec:problem}, the satellite tracking scenario considered in this work is illustrated in Fig.~\ref{fig:problem_setup}. This section develops the corresponding state-space model describing the satellite orbital dynamics. The nonlinear equations of motion are first presented and subsequently linearized around a nominal circular orbit to obtain a linear time-invariant representation suitable for Kalman filtering.

\subsection{State-Space Representation}
\label{sec:state_space}

Consider a satellite of unit mass moving under the influence of a central gravitational force. The motion is governed by Newton’s law under an inverse-square force field. Let $r(t)$ denote the radial distance of the satellite from the center of attraction and $\theta(t)$ denote its angular position at time $t$. The radial motion is described by
\begin{equation}
    \ddot{r}(t) = r(t)\dot{\theta}^2(t) - \frac{G}{r^2(t)}
    \label{eq:radial_dynamics}
\end{equation}
where $G$ represents the gravitational parameter of the system. The terms $\dot{r}(t)$ and $\ddot{r}(t)$ correspond to the radial velocity and radial acceleration, respectively, while $\dot{\theta}(t)$ denotes the angular velocity.

For an ideal circular orbit with constant radius $R$, the angular motion is uniform and can be expressed as $\theta(t)=\omega t$, where $\omega$ is the constant angular velocity. Under this condition, the gravitational parameter satisfies the relation $G = R^3\omega^2,$
and the angular dynamics are given by
\begin{equation}
    \ddot{\theta}(t) = -2 \dot{\theta}(t)\frac{\dot{r}(t)}{r(t)}
    \label{eq:angular_dynamics}
\end{equation}
To obtain a tractable estimation model, small deviations from the nominal circular orbit are assumed, namely, $\|x(t)\|\ll1$. The state vector is defined as
\begin{equation}
    x(t) = \left[x_1(t), x_2(t), x_3(t), x_4(t)\right]^\top \in \mathbb{R}^4
    \label{eq:state_vector}
\end{equation}
where the states represent deviations from the nominal motion. Specifically, $x_1(t)$ denotes the radial deviation from the nominal radius $R$, $x_2(t)$ represents the radial velocity, $x_3(t)$ corresponds to the scaled angular deviation, and $x_4(t)$ represents the scaled angular velocity deviation. These quantities are defined as
\begin{subequations}
\begin{align}
    x_1(t) &= r(t)-R,
    \label{eq:state_x1}\\
    x_2(t) &= \dot{r}(t),
    \label{eq:state_x2}\\
    x_3(t) &= R\big(\theta(t)-\omega t\big),
    \label{eq:state_x3}\\
    x_4(t) &= R\big(\dot{\theta}(t)-\omega\big).
    \label{eq:state_x4}
\end{align}
\end{subequations}

The scaling factor $R$ applied to the angular components ensures that all state variables are expressed in consistent physical units, which is advantageous for numerical stability in estimation. 
Applying a first-order Taylor expansion of \eqref{eq:radial_dynamics} and \eqref{eq:angular_dynamics} about the nominal orbit defined by $r(t)=R$ and $\theta(t)=\omega t$, and neglecting higher-order terms, yields the linearized continuous-time state-space model
\begin{equation}
    \dot{x}(t)
    =
    Ax(t),
    \label{eq:continuous_state}
\end{equation}
where $A\in\mathbb{R}^{4\times4}$ denotes the continuous-time system matrix obtained from the linearization of the nonlinear orbital dynamics.
For a sampling period $h>0$, the corresponding discrete-time model is
\begin{equation}
    x_{k+1}
    =
    Fx_k,
    \label{eq:discrete_state}
\end{equation}
where $F=e^{Ah}\in\mathbb{R}^{4\times4}$ denotes the discrete-time state-transition matrix obtained from the matrix exponential. Throughout this work, process disturbances are neglected, and uncertainty is introduced solely through the measurement model described in the following subsection.

\subsection{Measurement Model and Noise Characterization}
\label{sec:measurement_model}

Measurements are collected at discrete sampling times $t_k = kh$, where $k = 1, 2, \ldots, N$ and $h$ denotes the sampling interval. The observation model is defined as
\begin{equation}
    y_k = H x_k + v_k,
    \label{eq:measurement_model}
\end{equation}
where $y_k \in \mathbb{R}^2$ is the measurement vector, $x_k = x(t_k)$ is the state vector, and $v_k \in \mathbb{R}^2$ represents the measurement noise.
The measurement vector consists of the radial deviation and the scaled angular deviation, which can be written as
\begin{equation}
    y_k =
    \begin{bmatrix}
        y_1(t_k) \\
        y_3(t_k)
    \end{bmatrix}
    =
    \begin{bmatrix}
        x_1(t_k) \\
        x_3(t_k)
    \end{bmatrix}
    +
    \begin{bmatrix}
        v_1(t_k) \\
        v_3(t_k)
    \end{bmatrix}
    \label{eq:measurement_vector}
\end{equation}

The measurement noise $v_k = [v_1(t_k), v_3(t_k)]^\top$ is assumed to follow a zero-mean Gaussian distribution, 
$v_k\sim\mathcal{N}(0,\Sigma_v),$
where
\begin{equation}
    \Sigma_v=
    \begin{bmatrix}
        \varphi & 0\\
        0 & \psi
    \end{bmatrix}
    \label{eq:measurement_cov}
\end{equation}
is the measurement noise covariance matrix. The parameters $\varphi$ and $\psi$ denote the variances associated with the radial and angular measurements, respectively. Furthermore, the measurement noise components are assumed to satisfy
$\mathbb{E}[v_k]=0$ and $\mathbb{E}[v_kv_\ell^\top]=\Sigma_v\delta_{k\ell},$
where $\delta_{k\ell}$ denotes the Kronecker delta. Consequently, the measurement noise is mutually independent across sampling instants and uncorrelated with the system state.

\section{Kalman Estimation}
\label{sec:kalman}

Based on the mathematical model developed in Section~\ref{sec:model}, this section presents the state estimation framework. For completeness and to establish the notation used throughout the filtering algorithms, the discrete-time state-space model is recalled. The system is considered at discrete sampling instants $t_k=kh$, where $k=1,2,\ldots,N$ and $h$ denotes the sampling interval. The state-space model is given by
\begin{subequations}
\begin{align}
    x_k &= F x_{k-1} + q_{k-1}
    \label{eq:state_model} \\
    y_k &= H x_k + v_k
    \label{eq:measurement_model_kf}
\end{align}
\end{subequations}
where $x_k \in \mathbb{R}^n$ is the state vector, $y_k \in \mathbb{R}^m$ is the measurement vector, $F$ is the state transition matrix, and $H$ is the observation matrix. The process noise $q_k$ and measurement noise $v_k$ are assumed to be mutually independent, zero-mean Gaussian sequences with covariances $\Sigma_q$ and $\Sigma_v$, respectively.

The initial state $x_0$ is modeled as a Gaussian random variable with mean and covariance given by
\begin{equation}
    \mathbb{E}[x_0] = \hat{x}_{0|0}, \qquad
    \mathbb{E}\!\left[(x_0-\hat{x}_{0|0})(x_0-\hat{x}_{0|0})^\top\right]
    = P_{0|0}
    \label{eq:initial_state}
\end{equation}

The Kalman filter provides a recursive solution for state estimation through prediction and update steps. The one-step-ahead prediction is given by
\begin{subequations}
\begin{align}
    \hat{x}_{k|k-1} &= F \hat{x}_{k-1|k-1}
    \label{eq:state_prediction} \\
    P_{k|k-1} &= F P_{k-1|k-1} F^\top + \Sigma_q
    \label{eq:cov_prediction}
\end{align}
Upon receiving the measurement $y_k$, the estimate is updated as
\begin{align}
    K_k &= P_{k|k-1} H^\top \left(H P_{k|k-1} H^\top + \Sigma_v\right)^{-1}
    \label{eq:kalman_gain} \\
    \hat{x}_{k|k} &= \hat{x}_{k|k-1} + K_k \left(y_k - H \hat{x}_{k|k-1}\right)
    \label{eq:state_update} \\
    P_{k|k} &= \left(I - K_k H\right) P_{k|k-1}
    \label{eq:cov_update}
\end{align}
\end{subequations}
The term $y_k-H\hat{x}_{k|k-1}$ is referred to as the innovation, representing the discrepancy between the observed and predicted measurements. The Kalman gain $K_k$ is chosen to minimize the a posteriori estimation error covariance, yielding the minimum mean-square error (MMSE) linear estimator under the Gaussian assumptions introduced previously.

\subsection{Steady-State Gain Matrices}
\label{sec:steady_state}

The recursive implementation of the Kalman filter requires the computation of the error covariance matrix $P_{k|k-1}$ and the gain matrix $K_k$ at each time step. In particular, the evaluation of the Kalman gain involves the inversion of the innovation covariance matrix
\begin{equation}
    S_k = H P_{k|k-1} H^\top + \Sigma_v
    \label{eq:innovation_cov}
\end{equation}
which, in general, requires $O(n^3)$ floating-point operations for an $n\times n$ matrix. As a result, the computational complexity of the filter increases significantly with the dimension of the system, and repeated online evaluation of this matrix inversion may become computationally burdensome.

For time-invariant systems, it is therefore of practical interest to consider a steady-state formulation in which the covariance $P_{k|k-1}$ and gain $K_k$ matrices converge to constant values. In this case, the steady-state gain can be computed offline, thereby reducing the online computational burden while maintaining comparable estimation performance after the transient phase.

Starting from the covariance recursion
\begin{subequations}
\begin{align}
    P_{k|k-1} &= F P_{k-1|k-1} F^\top + \Sigma_q
    \label{eq:ss_cov_pred} \\
    P_{k|k} &= P_{k|k-1} - P_{k|k-1} H^\top
    \left(H P_{k|k-1} H^\top + \Sigma_v\right)^{-1}
    H P_{k|k-1}
    \label{eq:ss_cov_update}
\end{align}
it is of interest to determine whether the sequence $\{P_{k|k-1}\}$ converges as $k \to \infty$. If such a limit exists, then
\begin{equation}
    P_{k|k-1} \to P, \qquad K_k \to K
    \label{eq:ss_limit}
\end{equation}
\end{subequations}
for some constant matrices $P$ and $K$, since the Kalman gain $K_k$ depends explicitly on $P_{k|k-1}$.
In this case, the time-varying Kalman filter reduces to a steady-state estimator with constant gain.

By substituting the limiting covariance matrix into the Riccati recursion, the discrete algebraic Riccati equation (DARE) is obtained:
\begin{subequations}
\begin{equation}
    P = F \left(P - P H^\top \left(H P H^\top + \Sigma_v\right)^{-1} H P \right) F^\top + \Sigma_q
    \label{eq:dare}
\end{equation}
where $P$ represents the steady-state prediction error covariance matrix. Once the stabilizing solution \(P\) has been obtained, the corresponding steady-state Kalman gain is computed as
\begin{equation}
    K = P H^\top \left(H P H^\top + \Sigma_v\right)^{-1}.
    \label{eq:ss_gain}
\end{equation}
\end{subequations}
Using the steady-state gain, the Kalman filter reduces to the following predictor-corrector recursion:
\begin{subequations}
\begin{align}
    \hat{x}_{k+1|k} &= F \hat{x}_{k|k}
    \label{eq:ss_predictor} \\
    \hat{x}_{k|k} &= \hat{x}_{k|k-1} + K\left(y_k - H \hat{x}_{k|k-1}\right)
    \label{eq:ss_corrector}
\end{align}
\end{subequations}
where the innovation term $e_k = y_k - H \hat{x}_{k|k-1}$ represents the discrepancy between the actual measurement and its one-step-ahead prediction. 
Substituting \eqref{eq:ss_corrector} into \eqref{eq:ss_predictor} yields the equivalent compact predictor
\begin{subequations}
\begin{equation}
    \hat{x}_{k+1|k} = F \hat{x}_{k|k-1} + F K e_k.
    \label{eq:ss_compact_predictor}
\end{equation}
The associated predicted measurement is then
\begin{equation}
    \hat{y}_{k+1|k} = H \hat{x}_{k+1|k}.
    \label{eq:ss_output_predictor}
\end{equation}
\end{subequations}

The steady-state formulation is attractive because the gain matrix \(K\) is computed only once offline, thereby eliminating the repeated online solution of the Riccati recursion. This significantly reduces the computational burden for time-invariant systems while preserving nearly identical estimation performance after the transient response has decayed.

The existence and uniqueness of a stabilizing positive semidefinite solution to \eqref{eq:dare} depend on standard detectability and stabilizability conditions. In particular, if the pair $(F,H)$ is detectable and the pair $(F,\Sigma_q^{1/2})$ is stabilizable, then the Riccati recursion converges to the unique stabilizing solution of the DARE. 
Under these assumptions, the steady-state gain \(K\) guarantees that the estimation error dynamics are asymptotically stable. More precisely, the matrix
\begin{equation}
    F - K H
    \label{eq:error_dynamics_matrix}
\end{equation}
must be Schur stable, that is, all of its eigenvalues must lie strictly inside the unit circle. Consequently, the estimation error covariance remains bounded and the steady-state predictor is asymptotically stable.

\subsection{Micro-Kalman Filter ($\mu$KF)}
\label{sec:mukf}

The micro-Kalman filter ($\mu$KF) considered in this work is based on the information form of the Kalman filter, as introduced in \cite{b7,b8}. The formulation is developed for the discrete-time state-space model presented in Section~\ref{sec:kalman} and provides an alternative representation of the classical covariance-based filter. Rather than propagating the covariance matrix directly, the $\mu$KF operates on information quantities, leading to an equivalent estimation framework that is particularly attractive for distributed estimation.

In contrast to the classical Kalman filter, which propagates the covariance matrix $P_k$, the information form operates on the inverse covariance. Define the information matrix
\begin{equation}
    S_k = H^\top \Sigma_v^{-1} H
    \label{eq:info_matrix}
\end{equation}
and introduce the matrix
\begin{equation}
    M_k = \left(P_{k|k-1}^{-1} + S_k \right)^{-1}.
    \label{eq:M_matrix}
\end{equation}

The update step of the $\mu$KF can then be written as
\begin{subequations}
\begin{equation}
    \hat{x}_k = \hat{x}_{k|k-1} + M_k \left( H^\top \Sigma_v^{-1} y_k - S_k \hat{x}_{k|k-1} \right),
    \label{eq:mukf_update}
\end{equation}
which can be interpreted as an information-weighted correction of the prior estimate.

By defining the transformed measurement
\begin{equation}
    z_k = H^\top \Sigma_v^{-1} y_k,
    \label{eq:transformed_measurement}
\end{equation}
the update equation can be expressed more compactly as
\begin{equation}
    \hat{x}_k = \hat{x}_{k|k-1} + M_k \left( z_k - S_k \hat{x}_{k|k-1} \right).
    \label{eq:mukf_compact}
\end{equation}

Following the update, the prediction step is given by
\begin{align}
    \hat{x}_{k+1|k} &= F \hat{x}_k
    \label{eq:mukf_predict_state} \\
    P_{k+1|k} &= F M_k F^\top + \Sigma_q.
    \label{eq:mukf_predict_cov}
\end{align}
\end{subequations}

In this formulation, the matrix $M_k$ plays a role analogous to the posterior covariance in the classical Kalman filter. However, the update is expressed directly in terms of information quantities, which avoids explicit computation of the Kalman gain and enables efficient implementation in distributed settings.

Unlike the classical covariance formulation, the $\mu$KF performs the measurement update directly in the information domain before propagating the state estimate. Consequently, the estimator preserves the Bayesian interpretation of Kalman filtering while avoiding the explicit computation of the Kalman gain, making it particularly attractive for distributed estimation and sensor network applications.

\subsection{Neural-Scaled Micro-Kalman Filter}
\label{sec:neural_scaled_mukf}

The information-form $\mu$KF presented in the previous subsection assumes that the process and measurement noise covariance matrices, $\Sigma_q$ and $\Sigma_v$, are known and remain constant throughout the estimation process. In practice, however, the uncertainty statistics often vary with the operating conditions, causing the filter performance to deteriorate when the assumed covariance matrices no longer accurately represent the underlying system.

To improve the adaptability of the estimator while preserving the information-form filtering framework, a lightweight neural scaling mechanism is introduced. Instead of estimating the full covariance matrices directly, which generally requires estimating $\mathcal{O}(n^2)$ parameters while maintaining positive definiteness, the proposed approach estimates only two positive scaling factors. Consequently, the covariance structure is preserved while allowing the uncertainty level to vary online according to the observed measurements.

Let
\begin{equation}
    e_k
    =
    y_k-H\hat{x}_{k|k-1}
    \label{eq:innovation_nn}
\end{equation}
denote the innovation sequence. Since the innovation reflects the mismatch between the predicted and observed measurements, it naturally provides information regarding the current uncertainty of the estimation process. Accordingly, the neural network input is defined as
\begin{equation}
    \eta_k=
    \begin{bmatrix}
        \|e_k\|_2^2\\
        \|e_{k-1}\|_2^2\\
        1
    \end{bmatrix},
    \label{eq:nn_feature}
\end{equation}
where the constant entry provides a trainable bias.

The neural scaling factors are then computed according to
\begin{subequations}
\begin{align}
    \alpha_k
    &=
    \sigma
    \!\left(
    w_v^\top\eta_k
    \right),
    \label{eq:alpha_nn}\\
    \beta_k
    &=
    \sigma
    \!\left(
    w_q^\top\eta_k
    \right),
    \label{eq:beta_nn}
\end{align}
\end{subequations}
where $w_v$ and $w_q$ denote trainable weight vectors and $\sigma(\cdot)$ is a bounded activation function.

The nominal covariance matrices are subsequently adjusted according to
\begin{subequations}
\begin{align}
    \hat{\Sigma}_{v,k}
    &=
    \left(
    \alpha_{\min}
    +
    (\alpha_{\max}-\alpha_{\min})\alpha_k
    \right)\Sigma_v,
    \label{eq:scaled_measurement_cov}\\
    \hat{\Sigma}_{q,k}
    &=
    \left(
    \beta_{\min}
    +
    (\beta_{\max}-\beta_{\min})\beta_k
    \right)\Sigma_q,
    \label{eq:scaled_process_cov}
\end{align}
\end{subequations}
where the constants $\alpha_{\min}$, $\alpha_{\max}$, $\beta_{\min}$ and $\beta_{\max}$ define admissible scaling intervals. Since both scaling factors remain positive, the positive definiteness of the covariance matrices is preserved.

The adaptive information matrix is therefore given by
\begin{equation}
    \hat{S}_k
    =
    H^\top
    \hat{\Sigma}_{v,k}^{-1}
    H,
    \label{eq:scaled_information_matrix}
\end{equation}
and the information-form update becomes
\begin{subequations}
\begin{align}
    \hat{M}_k
    &=
    \left(
    P_{k|k-1}^{-1}
    +
    \hat{S}_k
    \right)^{-1},
    \label{eq:scaled_M}\\
    z_k
    &=
    H^\top
    \hat{\Sigma}_{v,k}^{-1}
    y_k,
    \label{eq:scaled_z}\\
    \hat{x}_{k|k}
    &=
    \hat{x}_{k|k-1}
    +
    \hat{M}_k
    \left(
    z_k-
    \hat{S}_k
    \hat{x}_{k|k-1}
    \right),
    \label{eq:scaled_update}
\end{align}
\end{subequations}
which retains exactly the same Bayesian information-form structure as the conventional $\mu$KF.

Finally, the prediction step is computed as
\begin{subequations}
\begin{align}
    \hat{x}_{k+1|k}
    &=
    F\hat{x}_{k|k},
    \label{eq:scaled_prediction_state}\\
    P_{k+1|k}
    &=
    F\hat{M}_kF^\top
    +
    \hat{\Sigma}_{q,k}.
    \label{eq:scaled_prediction_cov}
\end{align}
\end{subequations}

Consequently, the proposed Neural-Scaled $\mu$KF preserves the original information-form filtering equations while introducing an adaptive neural mechanism that continuously adjusts the uncertainty model according to the innovation sequence. The neural module therefore complements the $\mu$KF rather than replacing its Bayesian estimation framework.

\section{Simulation Results}

The simulation is initialized with the parameters $R = 1$, $\omega = 1$, and $G = 1$, and is evaluated over $N = 1000$ discrete time steps. The sampling interval is chosen as $h = 0.01$. 
The initial state is modeled as a Gaussian random variable
\begin{equation}
    x_0 \sim \mathcal{N}(\hat{x}_{0|0}, P_{0|0}),
    \label{eq:init_state}
\end{equation}
with mean $\hat{x}_{0|0} = [0.1,\, 0,\, 0,\, 0]^\top$ and covariance $P_{0|0} = 0.1 I_{4}$.

Based on the linearized dynamics, the continuous-time state equations are given by
\begin{align}\label{eq:dyn}
    \left.\begin{aligned}
    \dot{x}_1 &= x_2  \\
    \dot{x}_2 &= 3\omega^2 x_1 + 2\omega x_4 \\
    \dot{x}_3 &= x_4  \\
    \dot{x}_4 &= -2\omega x_2 
    \end{aligned}\quad\right\} \quad
    \dot{x}(t) = 
    \begin{bmatrix}
        0 & 1 & 0 & 0 \\
        3\omega^2 & 0 & 0 & 2\omega \\
        0 & 0 & 0 & 1 \\
        0 & -2\omega & 0 & 0
    \end{bmatrix} x(t),
\end{align}
Discretizing the system with sampling interval $h = 0.01$ yields the state transition matrix
\begin{equation}
    F =
    \begin{bmatrix}
        1.0001  & 0.0100  & 0 & 0.0001 \\
        0.0300  & 1.0000  & 0 & 0.0200 \\
        0        & -0.0001 & 1 & 0.0100 \\
        -0.0003 & -0.0200 & 0 & 0.9998
    \end{bmatrix}.
    \label{eq:F_matrix}
\end{equation}

The measurement model follows the formulation introduced in Section~III, where the observation matrix is given by
\begin{equation*}
    H =
    \begin{bmatrix}
        1 & 0 & 0 & 0 \\
        0 & 0 & 1 & 0
    \end{bmatrix}.
\end{equation*}
The measurement noise covariance is defined as $\Sigma_v = \mathrm{diag}(\varphi, \psi)$, with $\varphi = 0.1$ and $\psi = 0.5$, corresponding to radial and angular measurement uncertainties, respectively.

To quantitatively evaluate the estimation performance, the mean square estimation error (MSEE) is computed. Let $\beta_k=\|x_k-\hat{x}_k^{KF}\|_2$ denote the estimation error of the baseline Kalman filter, and let $\gamma_k=\|x_k-\hat{x}_k^{N\mu}\|_2$ denote the estimation error of the Neural-Scaled $\mu$KF at time step $k$. For each simulation run $j$, the MSEE is defined as
\begin{equation}
    \kappa_j = \frac{1}{N} \sum_{k=1}^{N} \beta_k^2,
    \qquad
    \Gamma_j = \frac{1}{N} \sum_{k=1}^{N} \gamma_k^2,
    \label{eq:msee}
\end{equation}
where $N$ is the total number of time steps. To reduce statistical variability, the simulations are repeated $\phi$ times, and the averaged MSEE is computed as
\begin{equation}
    \Xi_{\kappa} = \frac{1}{\phi} \sum_{j=1}^{\phi} \kappa_j,
    \qquad
    \Xi_{\Gamma} = \frac{1}{\phi} \sum_{j=1}^{\phi} \Gamma_j.
    \label{eq:avg_msee}
\end{equation}

\begin{table*}[t!]
    \caption{Mean Square Estimation Error (MSEE) $\Gamma_k$ of the proposed Neural-$\mu$KF over ten Monte Carlo simulations.}
    \centering
    \resizebox{0.85\textwidth}{!}{%
    \begin{tabular}{l | c c c c c c c c c c c}
    \hline
    \multirow{2}{*}{$x_{k+1|k}$} & \multicolumn{10}{c}{Iteration $\phi_i$} \\
    & 1 & 2 & 3 & 4 & 5 & 6 & 7 & 8 & 9 & 10 \\
    \hline
    $x_1$ & 0.0017 & 0.0018 & 0.0027 & 0.0019 & 0.0026 & 0.0014 & 0.0016 & 0.0028 & 0.0022 & 0.0027 \\
    $x_2$ & 0.0064 & 0.0044 & 0.0058 & 0.0035 & 0.0048 & 0.0035 & 0.0036 & 0.0042 & 0.0035 & 0.0061 \\
    $x_3$ & 0.0056 & 0.0056 & 0.0021 & 0.0019 & 0.0014 & 0.0027 & 0.0048 & 0.0043 & 0.0046 & 0.0070 \\
    $x_4$ & 0.0076 & 0.0029 & 0.0068 & 0.0044 & 0.0056 & 0.0033 & 0.0042 & 0.0055 & 0.0059 & 0.0050 \\
    \hline
    \end{tabular}%
    }
    \label{Table1}
\end{table*}

\begin{table}[t!]
	\captionsetup{font=small}
	\caption{AMSEE comparison of the baseline Kalman filter ($\Xi_{\kappa}$) and the proposed Neural-$\mu$KF ($\Xi_{\Gamma}$).}
	\centering
	\begin{tabular}{l|cc}
    \hline
    $x_{k+1|k}$ & $\Xi_{\kappa}$ & $\Xi_{\Gamma}$\\
	\hline
	$x_1$ & 0.0017 & 0.0021\\
	$x_2$ & 0.0048 & 0.0046\\
    $x_3$ & 0.0041 & 0.0040\\
    $x_4$ & 0.0036 & 0.0051\\
	\hline
	\end{tabular}
    \label{Table3}
\end{table}

\begin{figure*}[t!]
\centering
\subfloat[]{%
    \includegraphics[width=0.325\linewidth]{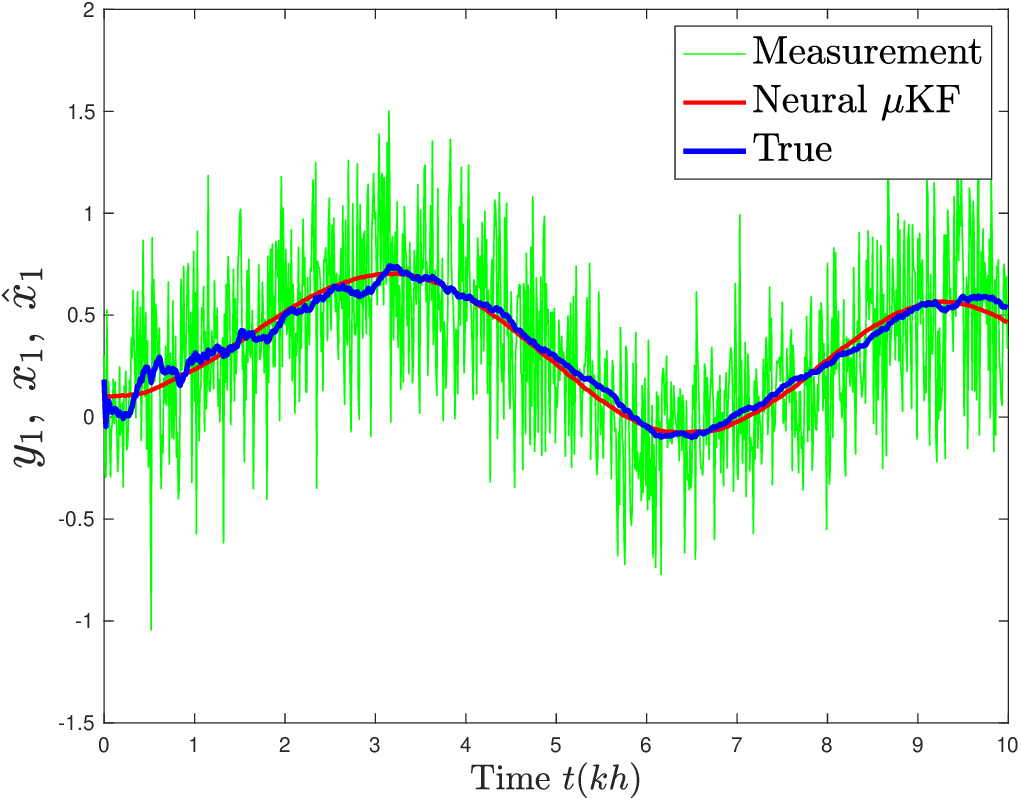}
    \label{fig:a1}
}
\subfloat[]{%
    \includegraphics[width=0.325\linewidth]{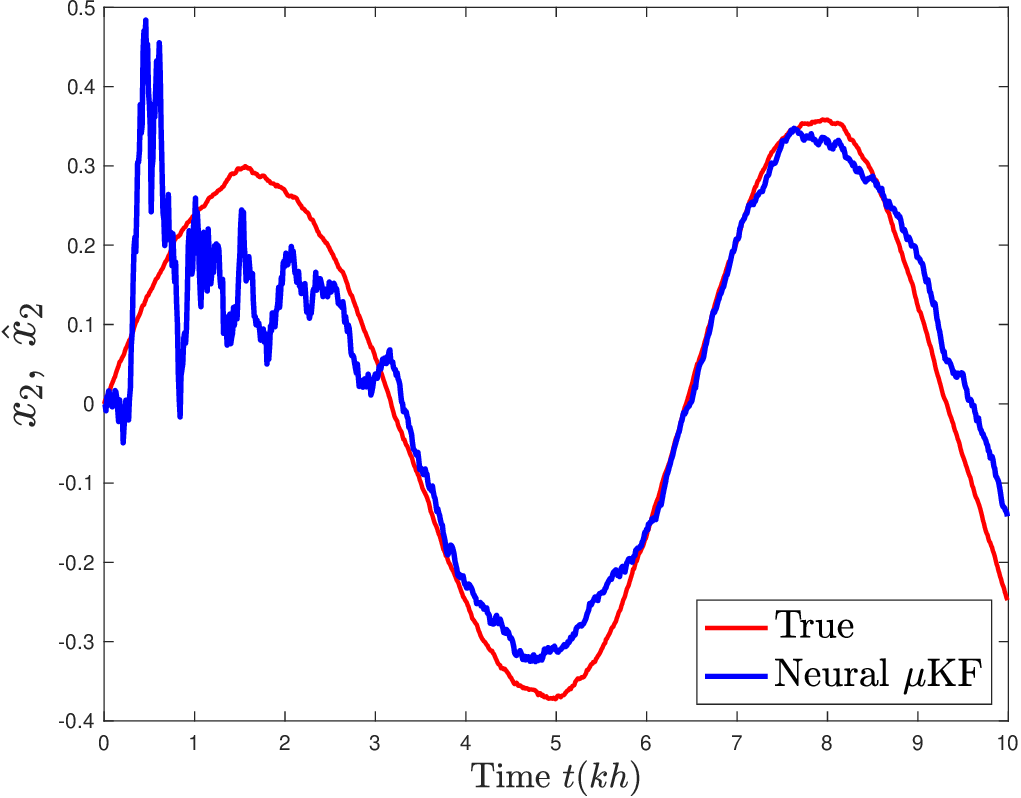}
    \label{fig:b1}
}
\subfloat[]{%
    \includegraphics[width=0.3195\linewidth]{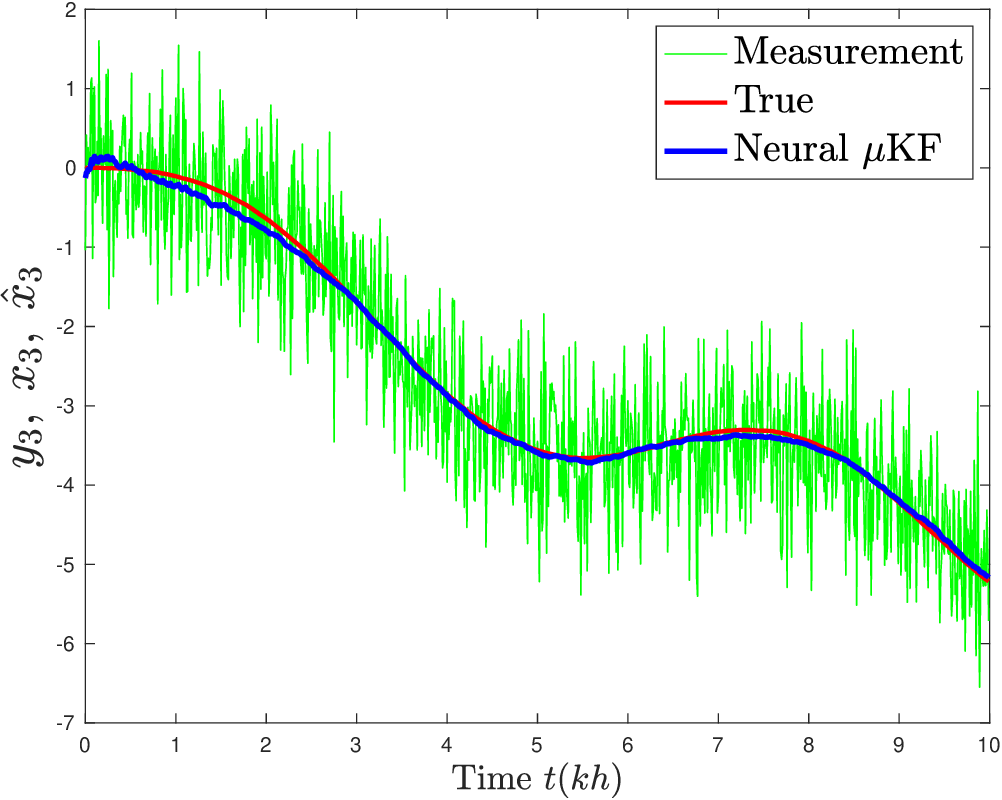}
    \label{fig:c1}
}

\medskip

\subfloat[]{%
    \includegraphics[width=0.325\linewidth]{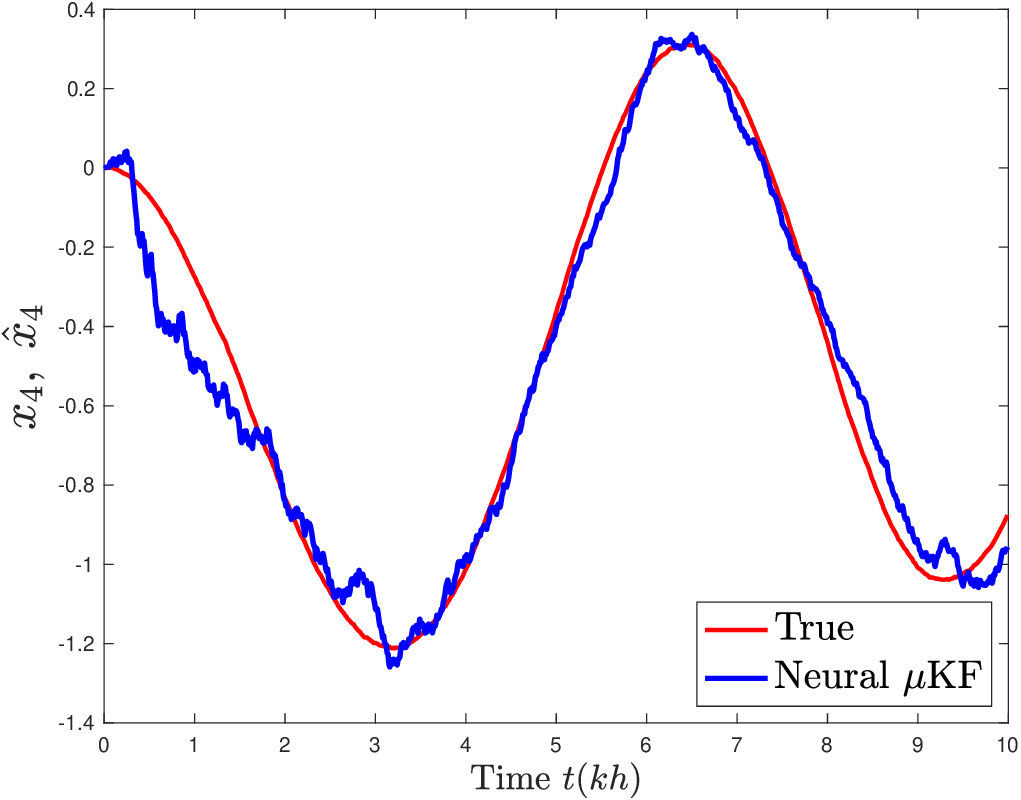}
    \label{fig:d1}
}\qquad
\subfloat[]{%
    \includegraphics[width=0.3375\linewidth]{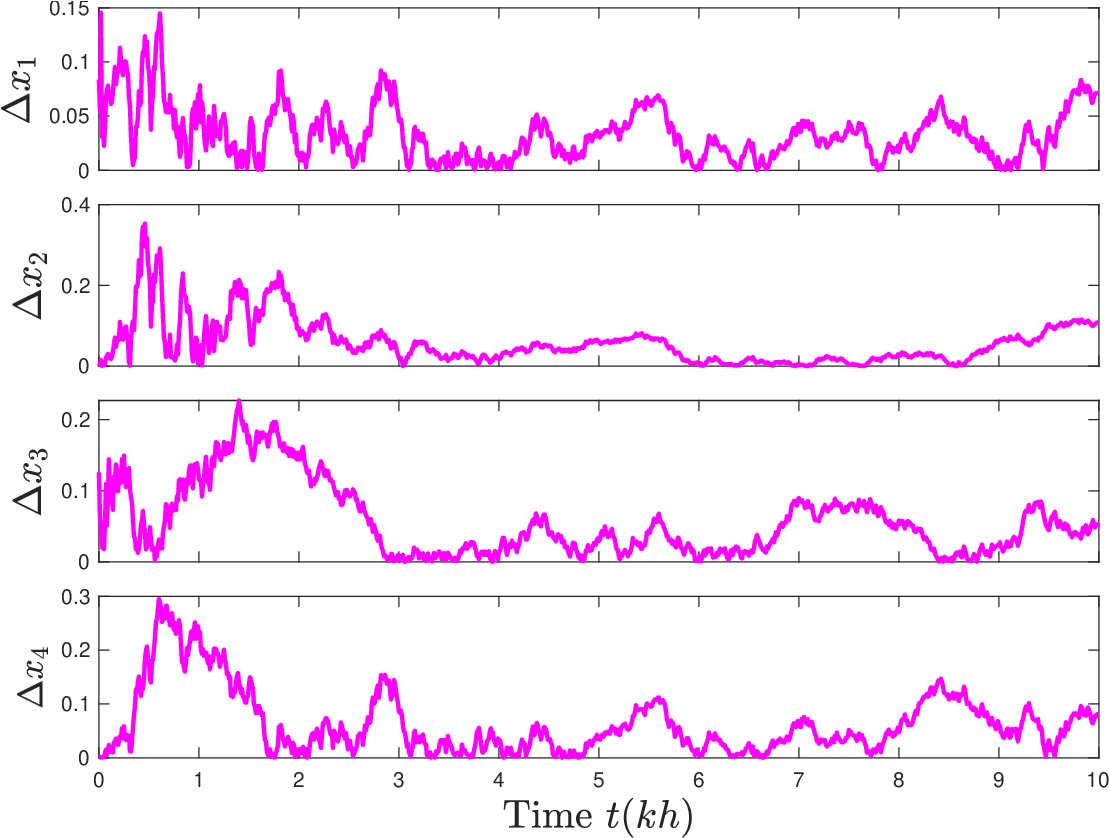}
    \label{fig:e1}
}
\caption{Filtering results obtained using the proposed Neural-$\mu$KF. Subplots (a)--(d) show the estimated and true trajectories for the radial position, radial velocity, scaled angular position, and scaled angular velocity states, respectively, while subplot (e) presents the corresponding estimation errors for all four states. The measurement information is included in subplots (a) and (c), corresponding to the measured states.}
\label{Fig:NeuralMuKF}
\end{figure*}

Figure~\ref{Fig:NeuralMuKF} illustrates the state estimation results obtained using the proposed Neural-$\mu$KF. Subplots~(a)--(d) compare the true and estimated trajectories of the radial position, radial velocity, scaled angular position, and scaled angular velocity, respectively, while subplot~(e) presents the corresponding estimation errors. The estimated trajectories closely follow the true system states throughout the simulation horizon, demonstrating the effectiveness of the proposed neural scaling strategy in preserving the estimation accuracy of the information-form filter.

The estimation performance is further quantified in Table~\ref{Table1}, which reports the Mean Square Estimation Error (MSEE) over ten independent Monte Carlo simulations. The results indicate that the proposed estimator consistently achieves small estimation errors for all state variables, with only minor variations across different simulation runs. This demonstrates the robustness of the Neural-$\mu$KF with respect to stochastic process and measurement noise.

To further assess the proposed approach, the averaged Mean Square Estimation Error (AMSEE) is compared with that of the baseline Kalman filter in Table~\ref{Table3}. The results show that the proposed Neural-$\mu$KF achieves a lower AMSEE for the radial velocity and scaled angular position states, while maintaining comparable performance for the remaining states. These results indicate that the neural covariance scaling mechanism can improve estimation accuracy without altering the underlying information-form filtering framework.

\begin{figure*}[t!]
\centering
\subfloat[]{%
    \includegraphics[width=0.325\linewidth]{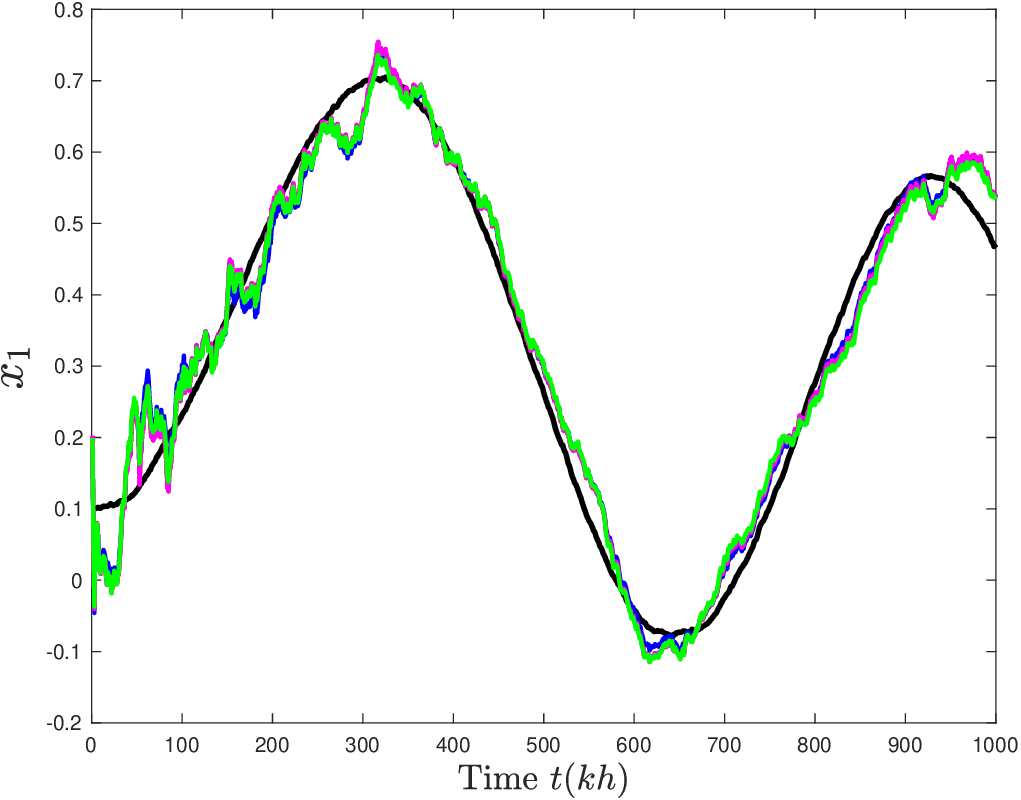}
    \label{fig:a3}
}
\subfloat[]{%
    \includegraphics[width=0.325\linewidth]{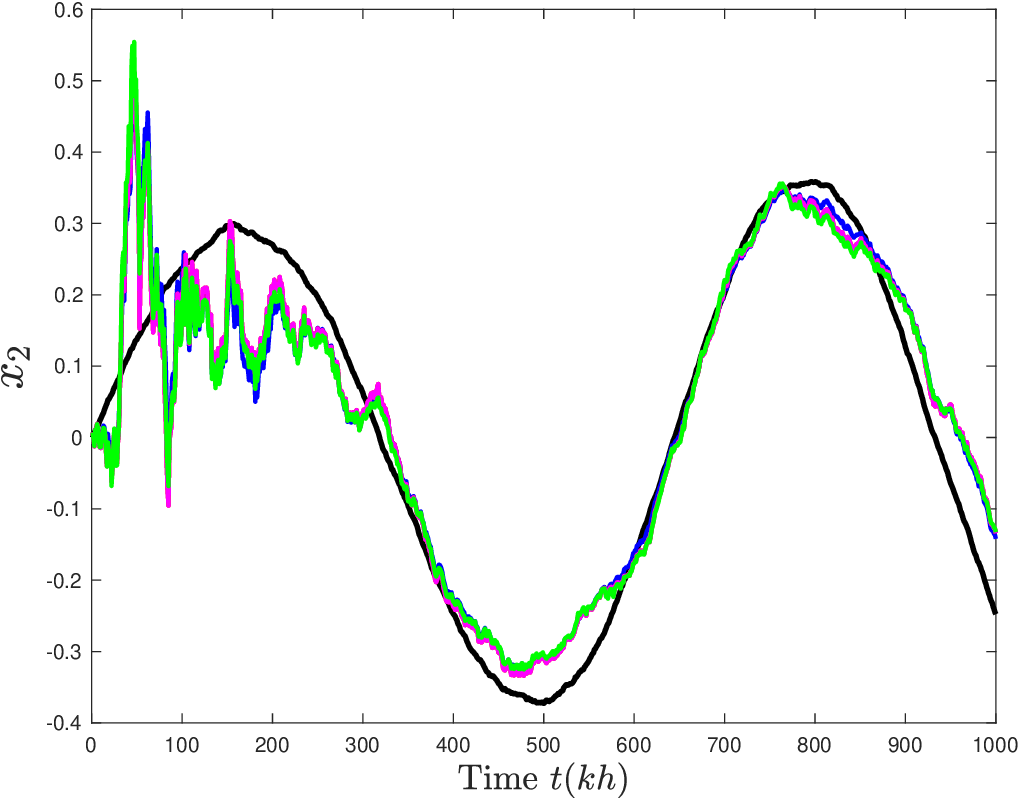}
    \label{fig:b3}
}
\subfloat[]{%
    \includegraphics[width=0.3195\linewidth]{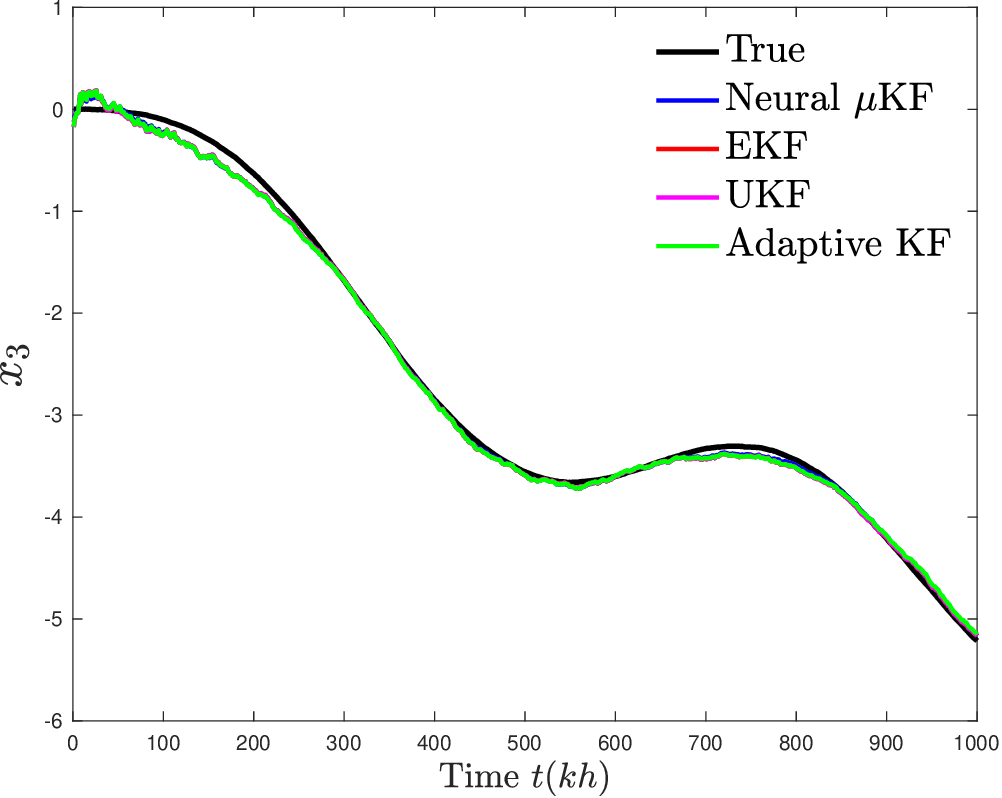}
    \label{fig:c3}
}

\medskip

\subfloat[]{%
    \includegraphics[width=0.325\linewidth]{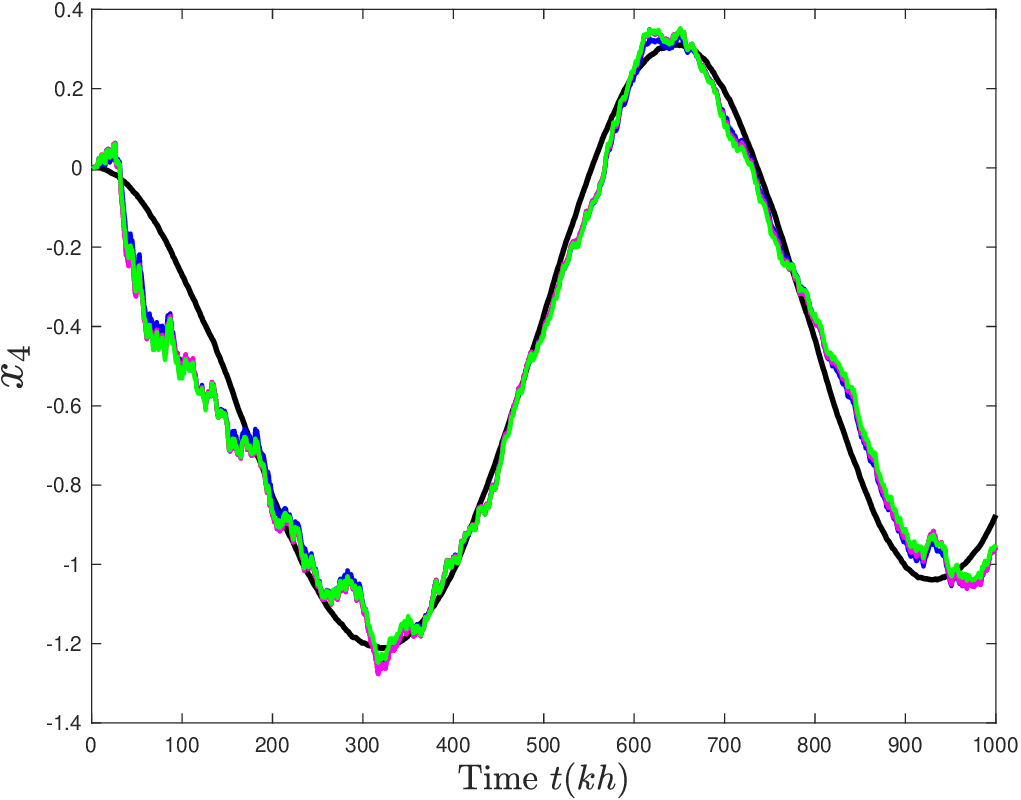}
    \label{fig:d3}
}\qquad
\subfloat[]{%
    \includegraphics[width=0.3375\linewidth]{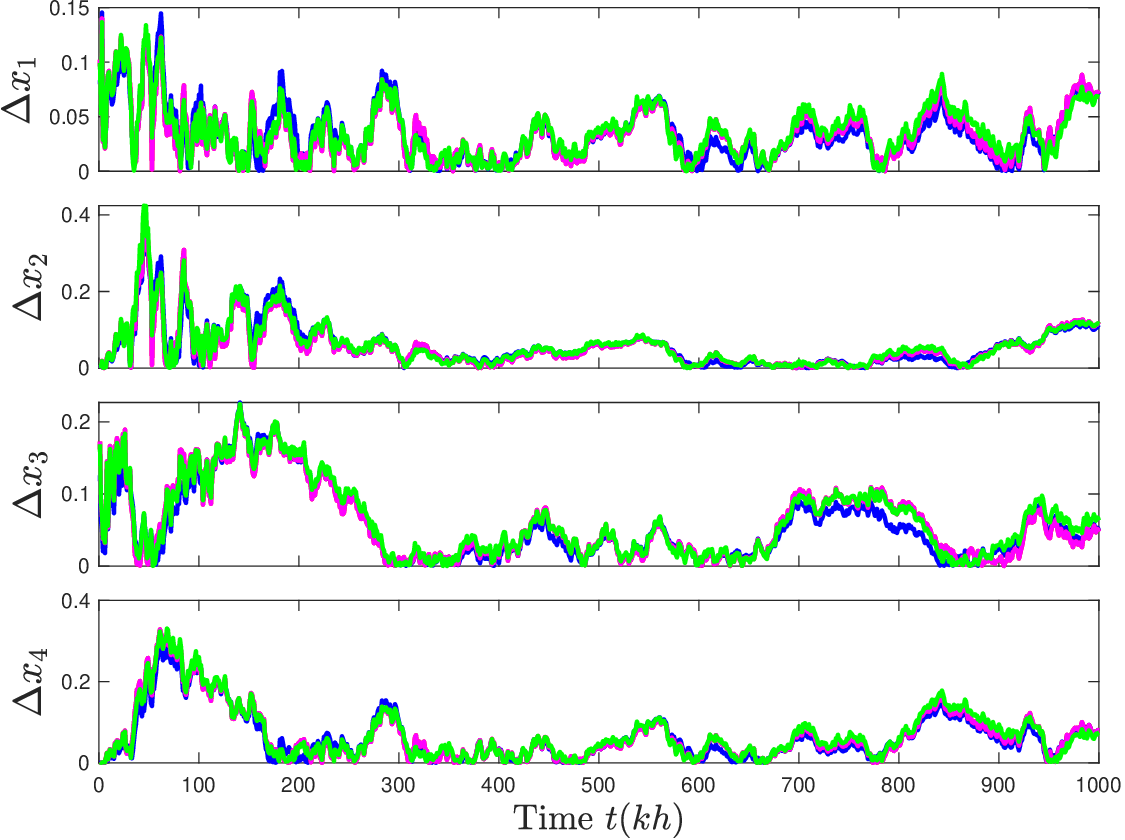}
    \label{fig:e3}
}
\caption{Comparison of the proposed Neural-$\mu$KF with the classical Kalman filter, EKF, UKF, and adaptive Kalman filter. Subplots (a)--(d) show the estimated and true trajectories for the radial position, radial velocity, scaled angular position, and scaled angular velocity states, respectively, while subplot (e) presents the corresponding estimation errors.}
\label{Fig:Comparison}
\end{figure*}

A broader comparison with the classical Kalman filter (KF), the Extended Kalman Filter (EKF), the Unscented Kalman Filter (UKF), and an adaptive Kalman filter is presented in Fig.~\ref{Fig:Comparison}. Subplots~(a)--(d) compare the estimated trajectories produced by each estimator, whereas subplot~(e) illustrates the corresponding estimation errors. As expected, all estimators accurately track the true states under the considered operating conditions. Owing to the linear nature of the satellite model, the EKF and UKF provide estimation performance similar to that of the classical KF, while the adaptive Kalman filter exhibits slight variations resulting from its online covariance adaptation mechanism.

Overall, the proposed Neural-$\mu$KF provides estimation performance that is comparable to, and in some states slightly better than, the conventional filtering approaches. These results demonstrate that incorporating a lightweight neural covariance scaling mechanism into the information-form $\mu$KF preserves the computational advantages of the original formulation while providing additional adaptability to measurement uncertainty.

\section{Conclusion}

This paper presented a Neural-enhanced information-form micro-Kalman filter ($\mu$KF) for satellite state estimation. Starting from a linearized state-space model of orbital dynamics, a lightweight neural scaling mechanism was incorporated into the information-form filtering framework to adapt the process and measurement noise covariances online while preserving the underlying Bayesian estimation structure.

The effectiveness of the proposed Neural-$\mu$KF was demonstrated through numerical simulations. The estimation results showed that the proposed method accurately tracked the satellite states with consistently low mean square estimation errors over multiple Monte Carlo simulations. Furthermore, comparison with the baseline Kalman filter indicated that the proposed approach achieved comparable, and in some states improved, estimation accuracy through adaptive covariance scaling.

A broader comparison with the extended Kalman filter (EKF), unscented Kalman filter (UKF), and adaptive Kalman filter further demonstrated that the proposed Neural-$\mu$KF provides estimation performance comparable to these established filtering methods under the considered satellite tracking problem. At the same time, it retains the computational advantages of the information-form formulation while introducing additional adaptability through a lightweight neural module.

Overall, the proposed Neural-$\mu$KF provides an effective and flexible state estimation framework for satellite tracking, combining the computational efficiency of information-based filtering with data-driven covariance adaptation. Future work will investigate the application of the proposed methodology to nonlinear orbital dynamics and more general aerospace estimation problems.

\section{Future Works}

Although the proposed Neural-$\mu$KF demonstrates promising performance for linear Gaussian satellite tracking, several research directions remain open.

A natural extension is the application of the proposed framework to nonlinear systems, where exact linearization, decoupling control, and unscented filtering become important for accurately describing nonlinear dynamics. These techniques have demonstrated encouraging results in process control and autonomous systems and may be extended to more realistic orbital models \cite{Wafi-ThreeTank,F1a,F1b,WAFI-JRC,F2a,F2b}.

Another direction is the development of more advanced adaptive estimation strategies. Future work may investigate online covariance adaptation, fault detection, and uncertainty quantification to improve robustness under unknown and time-varying noise conditions, particularly in hydraulic and industrial systems \cite{Wafi-Hydraulic,F3a,F3b}.

The information-form formulation adopted in this work also provides a natural basis for distributed state estimation. Extending the proposed Neural-$\mu$KF to decentralized and multi-agent systems represents a promising direction for large-scale interconnected applications \cite{Wafi-Quadruple,F4a,F4b}.

Finally, integrating the proposed estimator with advanced control strategies, including adaptive, robust, and distributed control, is another important research direction. Such integration may enable autonomous closed-loop systems capable of handling uncertainties, disturbances, delays, and communication constraints \cite{Wafi-Elham,F5a,F5b,Wafi-SIAM-arXiv,F6a,F6b,Wafi-MRAC,F7a,F7b}.

\section*{Acknowledgment}
Thanks to Professor Richard B. Vinter from the Imperial College London who has taught me in the lecture leading to finishing this paper and to LPDP (Indonesia Endowment Fund for Education) Scholarship from Indonesia.

\bibliographystyle{ieeetr}
\bibliography{Ref.bib}

\appendix
\section{Appendix}

This appendix summarizes the state estimation algorithms considered in this paper. For completeness, all filtering methods are formulated using the common discrete-time linear stochastic state-space model
\begin{align*}
    \begin{aligned}
    x_k &= Fx_{k-1}+q_{k-1},\\
    y_k &= Hx_k+v_k,
    \end{aligned}
\end{align*}
where $x_k\in\mathbb{R}^n$ is the system state, $y_k\in\mathbb{R}^m$ is the measurement vector, $F$ is the state-transition matrix, and $H$ is the observation matrix. The process and measurement noises are assumed to be mutually independent Gaussian random sequences satisfying
\[
    q_k\sim\mathcal{N}(0,\Sigma_q),
    \qquad
    v_k\sim\mathcal{N}(0,\Sigma_v).
\]
The corresponding Neural-$\mu$KF, EKF, and UKF algorithms are summarized below.

\begin{figure*}[h!]
\centering
\subfloat[]{%
    \includegraphics[width=0.33\linewidth]{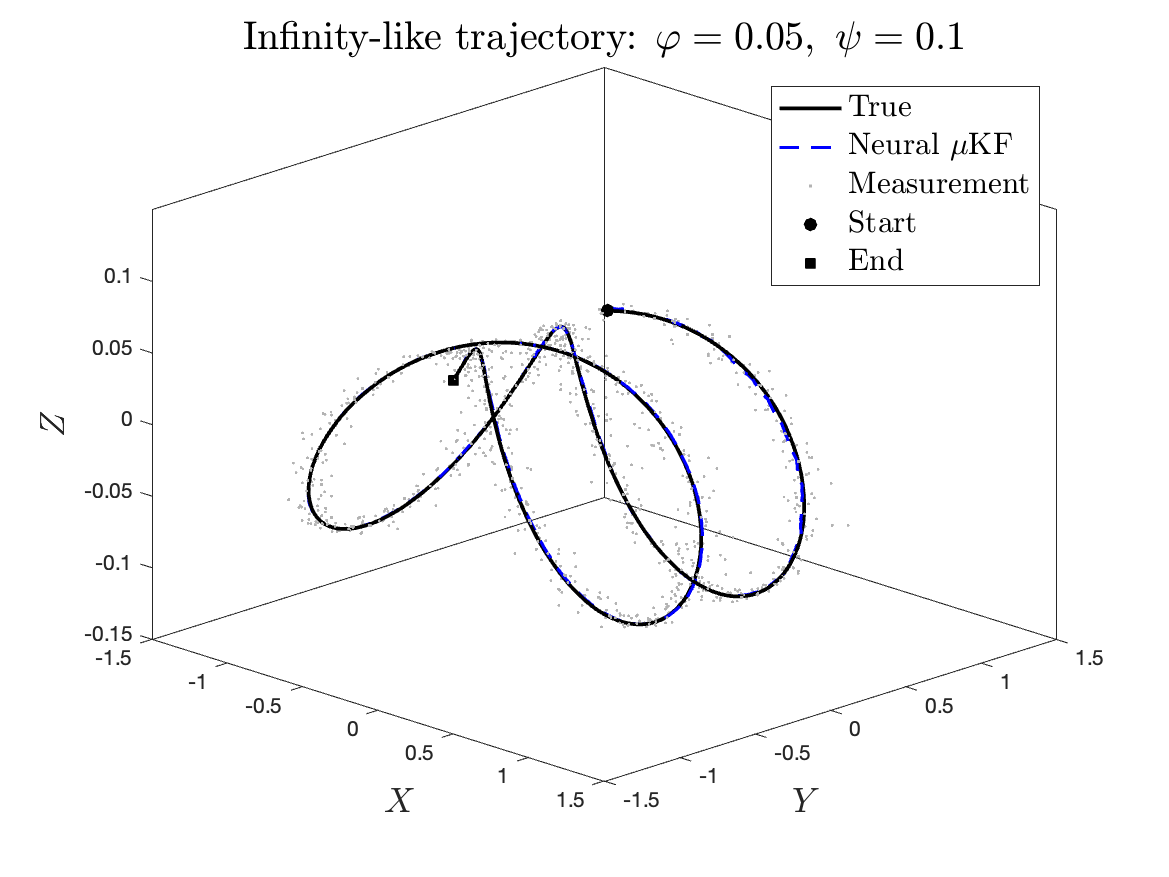}
    \label{fig:inf_a}
}
\subfloat[]{%
    \includegraphics[width=0.33\linewidth]{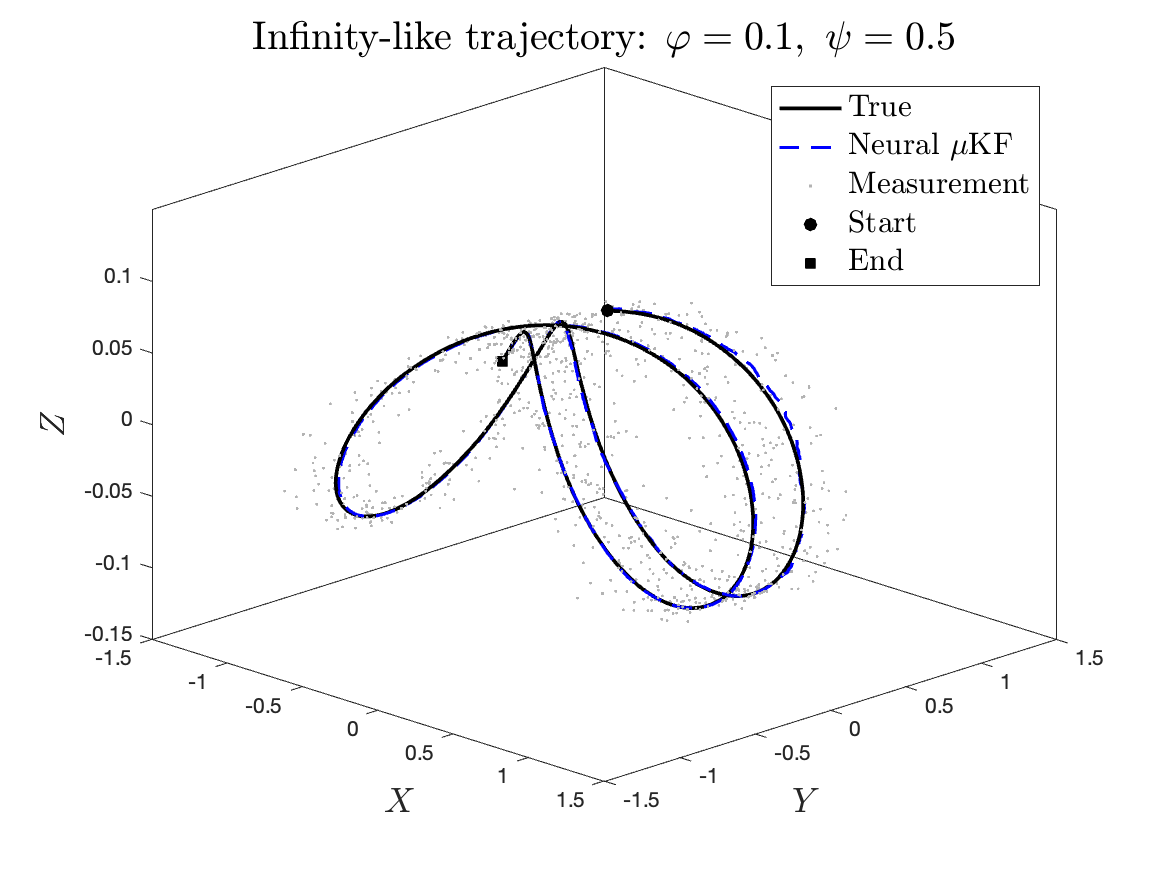}
    \label{fig:inf_b}
}
\subfloat[]{%
    \includegraphics[width=0.33\linewidth]{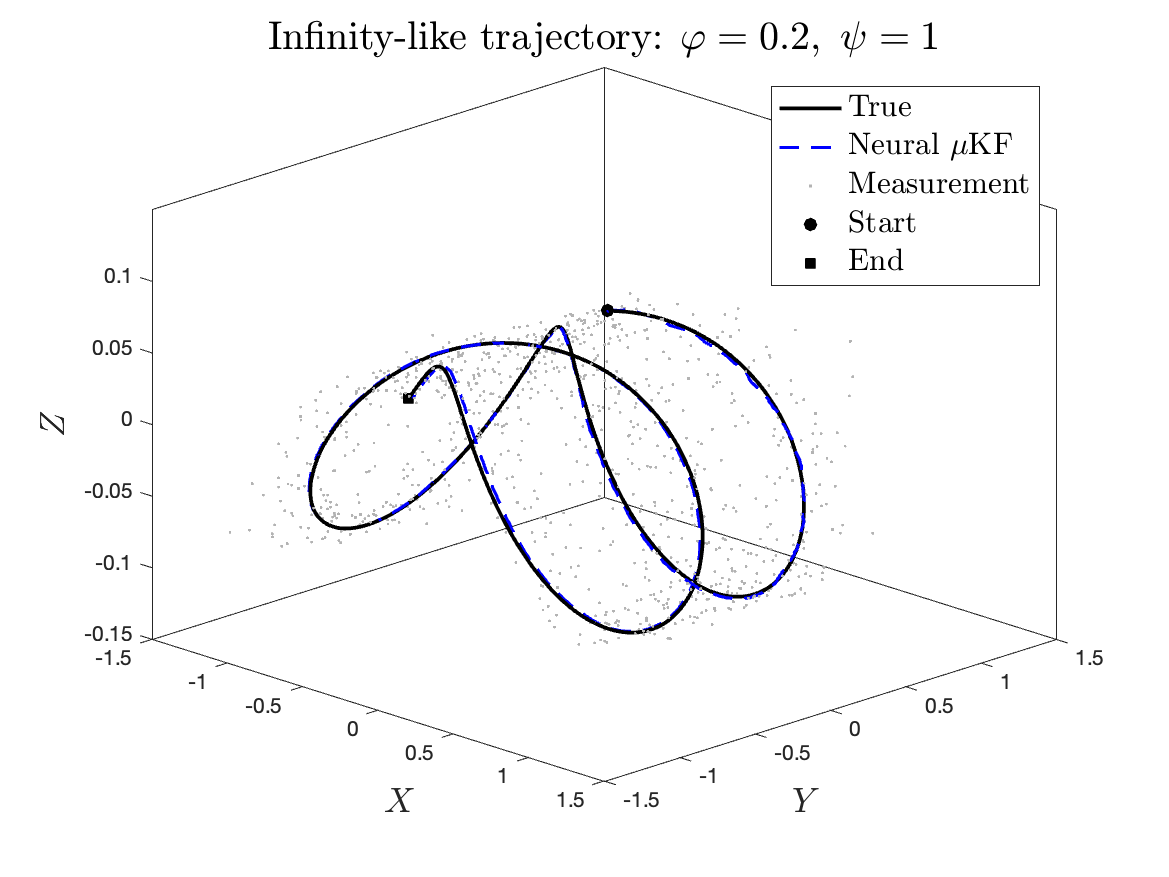}
    \label{fig:inf_c}
}
\caption{Infinity-shaped trajectory reconstruction using the proposed Neural-$\mu$KF under three measurement noise configurations: (a) low noise $(\varphi=0.05,\psi=0.10)$, (b) moderate noise $(\varphi=0.10,\psi=0.50)$, and (c) high noise $(\varphi=0.20,\psi=1.00)$. The true trajectory, noisy measurements, and the estimated trajectory are shown for comparison.}
\label{fig:infinity_compare}
\end{figure*}

\begin{algorithm}
\caption{Neural-Enhanced $\mu$-Kalman Filter}
\label{alg:neural_mukf}
\begin{algorithmic}[1]
\STATE Initialize $\hat{x}_{0|0}$, $P_{0|0}$, $\Sigma_q$, $\Sigma_v$
\FOR{$k=1,2,\ldots,N$}

    \STATE \textbf{Prediction:}
    \begin{align*}
    \hat{x}_{k|k-1} &= F\hat{x}_{k-1|k-1},\\
    P_{k|k-1} &= FP_{k-1|k-1}F^\top+\hat{\Sigma}_{q,k-1}.
    \end{align*}
    
    \STATE Compute the innovation
    \[
        e_k=y_k-H\hat{x}_{k|k-1}.
    \]
    
    \STATE Form the feature vector
    \[
        \eta_k=
        \begin{bmatrix}
        \|e_k\|_2^2\\
        \|e_{k-1}\|_2^2\\
        1
        \end{bmatrix}.
    \]
    
    \STATE Evaluate the neural scaling factors
    \begin{align*}
    \alpha_k = \sigma(w_v^\top\eta_k),\qquad
    \beta_k  = \sigma(w_q^\top\eta_k).
    \end{align*}
    
    \STATE Update the covariance matrices
    \begin{align*}
        \hat{\Sigma}_{v,k}
        &=
        \left(\alpha_{\min}
        +
        (\alpha_{\max}-\alpha_{\min})\alpha_k
        \right)\Sigma_v,\\
        \hat{\Sigma}_{q,k}
        &=
        \left(\beta_{\min}
        +
        (\beta_{\max}-\beta_{\min})\beta_k
        \right)\Sigma_q.
    \end{align*}
    
    \STATE \textbf{Information Update:}
    \begin{align*}
        \hat{S}_k &= H^\top\hat{\Sigma}_{v,k}^{-1}H,\\
        \hat{M}_k &= \left(P_{k|k-1}^{-1}+\hat{S}_k\right)^{-1},\\
        z_k &= H^\top\hat{\Sigma}_{v,k}^{-1}y_k.
    \end{align*}
    
    \STATE \textbf{State Update:}
    \[
        \hat{x}_{k|k}
        =
        \hat{x}_{k|k-1}
        +
        \hat{M}_k
        \left(
        z_k-\hat{S}_k\hat{x}_{k|k-1}
        \right).
    \]
\ENDFOR
\end{algorithmic}
\end{algorithm}

\begin{algorithm}
\caption{Extended Kalman Filter (EKF)}
\begin{algorithmic}[1]
\STATE Initialize $\hat{x}_{0|0}, P_{0|0}$
\FOR{$k = 1,2,\dots,N$}
    \STATE \textbf{Prediction:}
    \begin{align*}
        \hat{x}_{k|k-1} &= f(\hat{x}_{k-1|k-1}), \qquad 
        F_k = \left.\frac{\partial f}{\partial x}\right|_{\hat{x}_{k-1|k-1}} \\
        P_{k|k-1} &= F_k P_{k-1|k-1} F_k^\top + \Sigma_q
    \end{align*}
    \STATE \textbf{Update:}
    \begin{align*}
        H_k &= \left.\frac{\partial h}{\partial x}\right|_{\hat{x}_{k|k-1}} \\
        K_k &= P_{k|k-1} H_k^\top (H_k P_{k|k-1} H_k^\top + \Sigma_v)^{-1} \\
        \hat{x}_{k|k} &= \hat{x}_{k|k-1} + K_k (y_k - h(\hat{x}_{k|k-1})) \\
        P_{k|k} &= (I - K_k H_k) P_{k|k-1}
    \end{align*}
\ENDFOR
\end{algorithmic}
\end{algorithm}

\begin{algorithm}
\caption{Unscented Kalman Filter (UKF)}
\begin{algorithmic}[1]
\STATE Initialize $\hat{x}_{0|0}, P_{0|0}$
\FOR{$k = 1,2,\dots,N$}
    \STATE \textbf{Sigma Points:}
    \begin{align*}
        \chi_{k-1}^{(i)} = \text{SigmaPoints}(\hat{x}_{k-1|k-1}, P_{k-1|k-1})
    \end{align*}
    \STATE \textbf{Prediction:}
    \begin{align*}
        \chi_k^{(i)} &= f(\chi_{k-1}^{(i)}) \\
        \hat{x}_{k|k-1} &= \sum_i W_i \chi_k^{(i)} \\
        P_{k|k-1} &= \sum_i W_i (\chi_k^{(i)} - \hat{x}_{k|k-1})(\cdot)^\top + \Sigma_q
    \end{align*}
    \STATE \textbf{Measurement Prediction:}
    \begin{align*}
        y_k^{(i)} &= h(\chi_k^{(i)}) \quad \longrightarrow \quad
        \hat{y}_k = \sum_i W_i y_k^{(i)}
    \end{align*}
    \STATE \textbf{Update:}
    \begin{align*}
        P_{yy} &= \sum_i W_i (y_k^{(i)} - \hat{y}_k)(\cdot)^\top + \Sigma_v, \qquad 
        P_{xy} = \sum_i W_i (\chi_k^{(i)} - \hat{x}_{k|k-1})(y_k^{(i)} - \hat{y}_k)^\top \\
        K_k &= P_{xy} P_{yy}^{-1} \\
        \hat{x}_{k|k} &= \hat{x}_{k|k-1} + K_k (y_k - \hat{y}_k) \\
        P_{k|k} &= P_{k|k-1} - K_k P_{yy} K_k^\top
    \end{align*}
\ENDFOR
\end{algorithmic}
\end{algorithm}


\begin{algorithm}
\caption{Adaptive Kalman Filter (Covariance Estimation)}
\begin{algorithmic}[1]
\STATE Initialize $\hat{x}_{0|0}, P_{0|0}, \Sigma_q, \Sigma_v$
\FOR{$k = 1,2,\dots,N$}
    \STATE \textbf{Prediction:}
    \begin{align*}
        \hat{x}_{k|k-1} &= F \hat{x}_{k-1|k-1} \\
        P_{k|k-1} &= F P_{k-1|k-1} F^\top + \Sigma_q
    \end{align*}
    \STATE \textbf{Innovation:}
    \begin{align*}
        e_k = y_k - H \hat{x}_{k|k-1}
    \end{align*}
    \STATE \textbf{Adaptive Update:}
    \begin{align*}
        \Sigma_v &= \alpha \Sigma_v + (1-\alpha) (e_k e_k^\top)
    \end{align*}
    \STATE \textbf{Kalman Update:}
    \begin{align*}
        K_k &= P_{k|k-1} H^\top (H P_{k|k-1} H^\top + \Sigma_v)^{-1} \\
        \hat{x}_{k|k} &= \hat{x}_{k|k-1} + K_k e_k \\
        P_{k|k} &= (I - K_k H) P_{k|k-1}
    \end{align*}
\ENDFOR
\end{algorithmic}
\end{algorithm}

\end{document}